\documentclass[10pt,a4paper,twoside]{article}

\usepackage{vmargin,fancyheadings,multicol,multirow,ifthen,cite,
    epsfig,wrapfig,calc,dcolumn,apalike,
    setspace, 
    boxedminipage,rotating,
    textcomp, 
        booktabs, 
    picinpar,longtable}

\usepackage{imc2024}

\begin{document}
\SetPaperBodyFont   



\setcounter{page}{1}

\begin{IMCpaper}{

\title{Atmospheric height measurement of meteors detected simultaneously
by Digisonde and multi-site optical cameras during the 2019 Geminid shower}

\author{Lívia Deme$^1$,
Csilla Szárnya$^{1,2}$
Veronika Barta$^1$,
Antal Igaz$^3$,
Krisztián Sárneczky$^3$,
Balázs Csák$^3$,
Nándor Opitz$^3$,
Nóra Egei$^3$,
József Vinkó$^3$
}

\abstract{Simultaneous optical and ionosonde detections of meteors offer 
a great opportunity to measure the transient physical properties of the 
meteor's ionization trail. One of the key parameters of the ionization 
trail is its true geometric height above the surface of the Earth, which 
can be determined from the trajectory of the optical meteors by multi-site 
measurements. During the peak of the 2019 Geminid shower we looked for 
optical meteors among the Konkoly Meteor Observatory Network data taken 
at ELTE Gothard Observatory, Szombathely, and contemporaneous meteor 
signals detected by the DPS-4D type ionosonde (Digisonde) operating at 
Széchenyi István Geophysical Observatory (SZIGO), Nagycenk, Hungary. 
From the 2 simultaneous detections we inferred the atmospheric height 
of the meteors using the method of intersecting planes. The results, 
about 90 and 85 km, are consistent with the detection limit (above 80 km) 
of the Digisonde.}
\datereceived{Received 2024 Oct 31} 
}

\section{Introduction}
\thankstext{HUN-REN Institute of Earth Physics and Space Science, Sopron, Hungary}
\thankstext{Doctoral School of Earth Sciences, ELTE Eötvös Loránd University, Budapest, Hungary}
\thankstext{Konkoly Observatory, HUN-REN Institute for Astronomy and Earth Sciences, Budapest, Hungary}

    \setcounter{footnote}{0} 

Meteors can create ionization trails along their trajectory in the lower part of Earth's ionosphere (the
so-called D and E-region), which is an under-explored phenomenon in the upper atmosphere. 
Such ionization trails can survive for minutes in certain cases,
depending on local physical conditions such as magnetic field and wind velocity. These
represent rapid transient disturbances in the ionosphere, and they can be detected by
measuring the reflected radio impulses emitted upward by an ionosphere radar, called
ionosonde (\cite{Maruyama2003}).  

When meteorites enter the atmosphere and burn up at an altitude of $\sim$70-110 km, dust and metallic material is
deposited in the upper atmosphere. At night, these can absorb electrons, changing the balance between ions and
electrons (\cite{Friedrich2012}). Interesting variations have also been observed in the legendary ionospheric
phenomenon known as sporadic E layer (which, according to current theories, is meteoric metallic dust collected by
the Earth's gravitational field and trapped in a thin layer by the upper atmosphere's winds) during major meteor
showers (\cite{Haldoupis2011,Chandra2001,Jacobi2013}).

Burning meteors in the upper atmosphere can be approximated by a metallic cylinder with a length in the $\sim$10 km range, and a radius in the 0.5-4 m range that is height dependent (\cite{Stuart1970}). The ionization trail is
affected by amplitude diffusion, the wind shear and turbulence of local winds (\cite{Kozlovsky2018,Maruyama2003}).
According to \cite{McKinley1961}, the brightness of the meteor affects where the ionization trace appears: bright meteors (below 0th magnitude) have the ionization trace above the height of the light-emitting trail, while it is the opposite for fainter meteors. 

Meteor characteristics are also strongly influenced by the meteor's initial physical parameters, such as mass and
entry velocity. The higher the velocity for a given mass, the higher the altitude of the meteor trail. 
On the other hand, at a given speed the higher the mass, the lower the altitude of the trails (\cite{Stuart1970}).
The brighter the meteor, the higher the electron line density (electrons per meter of trail length): roughly $10^{15}$ el/m is associated with a visual magnitude of $2.5$, while $-2.5$ is associated with $10^{17}$ el/m (\cite{Manning1959}).

The ionosphere, an environment with changing electron concentration, can refract or reflect an electromagnetic 
(EM) wave propagating into it (\cite{davies90}). 
If the EM wave comes from the Earth's surface (emitted by the ionosonde), 
it is reflected back if its frequency is lower than the so-called plasma frequency:

\begin{equation}
f_p = {\omega_{pe} \over {2 \pi}} ~=~ {1 \over {2 \pi}} \sqrt{ {n_e c^2 } \over {m^* \varepsilon_0} },
\label{eq:1}
\end{equation}
where $f_p$ is the electron plasma frequency (in Hz), $\omega_{pe}$ is the electron plasma angular 
frequency (in radian/s), $n_e$ is the electron number density (in 1/m$^3$), $e$ is the electron's charge,
$m^*  = 9.11 \times 10^{-31}$ kg is the electron's effective mass and $\varepsilon_0 = 8.85 \times 10^{-12}$ 
F/m is the dielectric constant (vacuum permittivity). 
This creates a possibility for scanning 
different layers of the ionosphere with an ionosonde by emitting EM waves with different frequencies 
and measuring the reflected signal. When the frequency of the wave ($f$) reaches $f_p$ at a certain
layer, (which is called the critical frequency of the layer) the EM wave is absorbed by the ionosphere
(in other words, its time-of-flight becomes infinite). If $f > f_p$, there will be no reflected signal
as the EM wave goes through the medium.  

\cite{Maruyama2003} demonstrated that ionosondes (10C type) can detect the ionization traces of individual
meteoroides. The specification of the instruments may strongly influence how many events they can detect 
(\cite{zbysek24}).

The modern version of the ionosonde, called Digisonde (DPS-4D type ionosonde),
emits impulses in radio frequencies, typically from 1 to 20 MHz, and measures the signal
reflected by different layers in the ionosphere. This instrument has extra features: an antenna system 
capable of direction measurement (although with lower angular resolution), and it also has a mode of operation that is suitable for monitoring drift motion of the plasma (\cite{Reinisch2011}). 
\cite{Kereszturi2021} and \cite{szarnya2023} proved that it is possible to detect individual meteoroid 
bodies with a Digisonde.

The Digisonde measures the height of a reflecting source via the time delay between the emitted and reflected
signal (called ``time-of-flight'' measurement). Local enhancements of the
ionization, such as meteor trails, create a transient signal on the ionogram, where the height of
the echo (in kilometers) is plotted against the radio frequency (in MHz). 

The problem with this kind of measurement is the geometric distortion of the flight path for 
off-zenith sour{\-}ces. Since the Digisonde measures the time-of-flight of the signal, it
assigns higher distances to sources at lower elevation. Typically, meteors with elevation of at
least 40 degrees or higher can be detected by a Digisonde (\cite{Kozlovsky2018,szarnya24}).

Another issue with the interpretation of the Digisonde's raw measurements is the initial assumption 
that the medium,
in which the EM signal propagates between the Earth's surface and the ionosphere,
is assumed to be vacuum.  This can be taken into account with inversion models, but it is not 
applied automatically.
Together with the geometric distortion, these two effects virtually increase the flight path 
(computed with the assumption that the reflecting source is at the zenith and the signal propagates in 
vacuum). 

Meteors can provide a unique and important possibility for determining the actual atmospheric
height of their trajectories, and constrain the virtual height measurements made by the Digisonde. The
Digisonde also provides information on the ionization caused by meteoroids. This was one of
the main purposes of the campaign that was performed at Sopron Digisonde station during the
Geminid shower in 2019. The main results of that campaign is summarized in \cite{szarnya24}: it was found
that there is an anti-correlation between meteor brightness and the maximum frequency of the traces 
detected by the Digisonde, while the meteor velocity and the maximum frequency of ionization are positively
correlated. 

\begin{figure*}
\epsfig{file=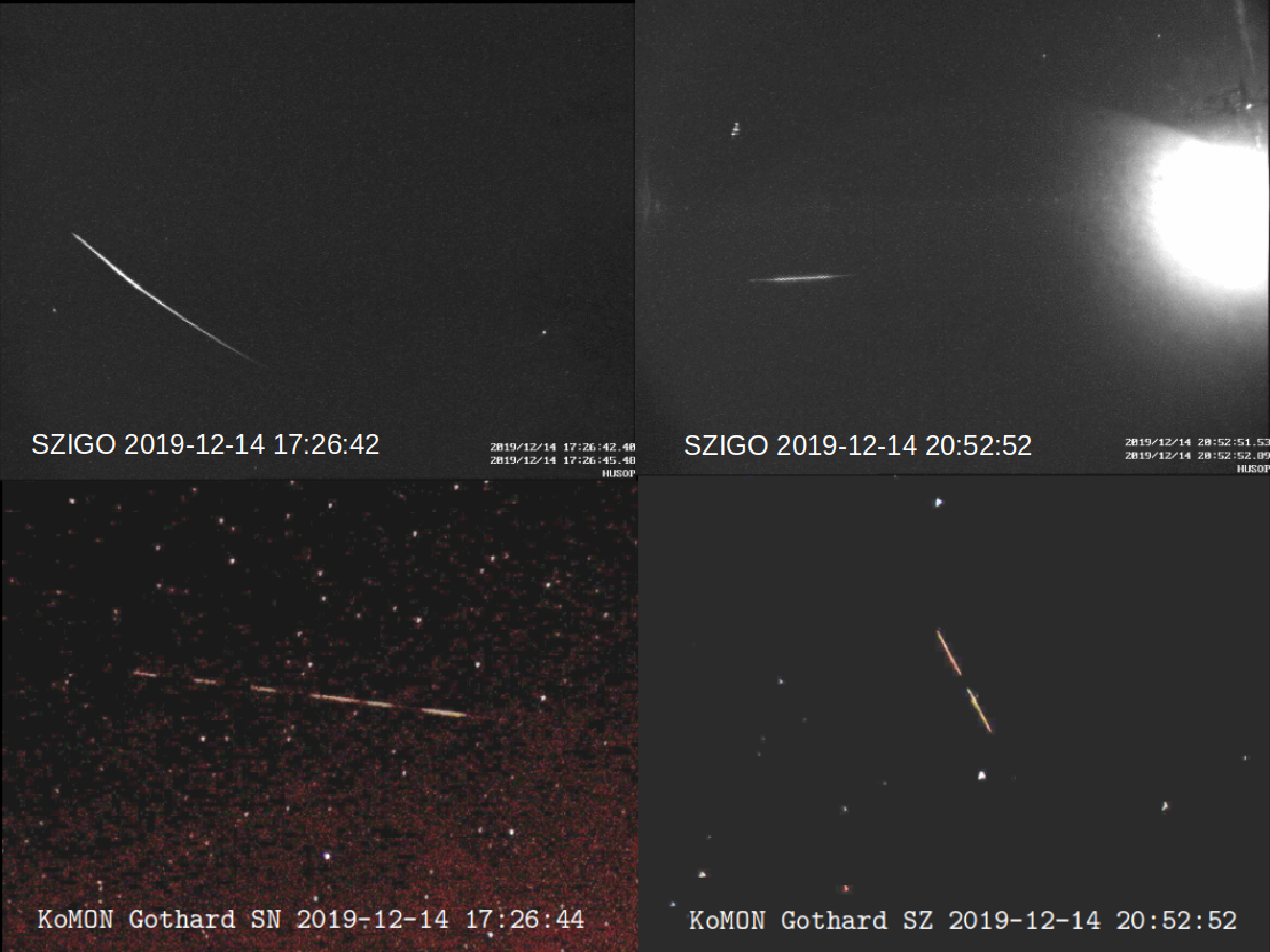,width=\textwidth}
\caption{\label{fig:1} optical images of the two simultaneous meteors detected by the Digisonde}
\end{figure*}

\begin{table*}
\caption{Basic properties of the two simultaneous meteors detected on December 14, 2019. The moment of first detection (in UT) and the horizontal coordinates (Altitude and Azimuth) at the beginning and the end of the trajectory are shown.}
\vspace{0mm}
\begin{center}
\begin{tabular}{|l|c|c|c|c|c|} \hline
Meteor No. & UT & ALT(beg) & AZ(beg) & ALT(end) & AZ(end)\\
\hline \hline
SZIGO No. 1 & 17:26:42 & 54$^\circ$.03 & 325$^\circ$.55 & 38$^\circ$.92 & 79$^\circ$.87 \\
Gothard No. 1 & 17:26:44 & 36$^\circ$.86 & 303$^\circ$.35 & 39$^\circ$.72 & 314$^\circ$.62 \\
SZIGO No. 2 & 20:52:52 & 57$^\circ$.88 & 305$^\circ$.16 & 47$^\circ$.44 & 298$^\circ$.23 \\
Gothard No. 2 & 20:52:52 & 39$^\circ$.46 & 325$^\circ$.95 & 41$^\circ$.83 & 328$^\circ$.56 \\
\hline
\end{tabular}
\end{center}
\label{tab:1}
\end{table*}

\begin{table*}
\caption{Calculated heights (in kilometers) of the two simultaneous meteors. $H_{beg}$ and $H_{end}$ denotes 
the height at the beginning and end of the computed meteor trajectory at each site. 
$H_{virt}$ is the virtual height measured by the Digisonde and $H_{corr}$ is the Digisonde height 
corrected for the meteor zenith distance at the first detection. The uncertainty of the Digisonde height
measurements is about $\pm 2.5$ km.}
\vspace{0mm}
\begin{center}
\begin{tabular}{|l|c|c|c|c|c|c|} \hline
UT & SZIGO $H_{beg}$ & SZIGO $H_{end}$ & Gothard $H_{beg}$ & Gothard $H_{end}$ & $H_{virt}$ & $H_{corr}$ \\
(hh:mm:ss) & (km) & (km) & (km) & (km) & (km) & (km) \\
\hline \hline
17:26:42 & 89.9 & 82.7 & 84.8 & 82.7 & 114.8 & 92.9 \\
20:52:51 & 85.4 & 73.6 & 78.0 & 74.2 & 96.7 & 81.9 \\
\hline
\end{tabular}
\end{center}
\label{tab:2}
\end{table*}

\begin{figure*}
\epsfig{file=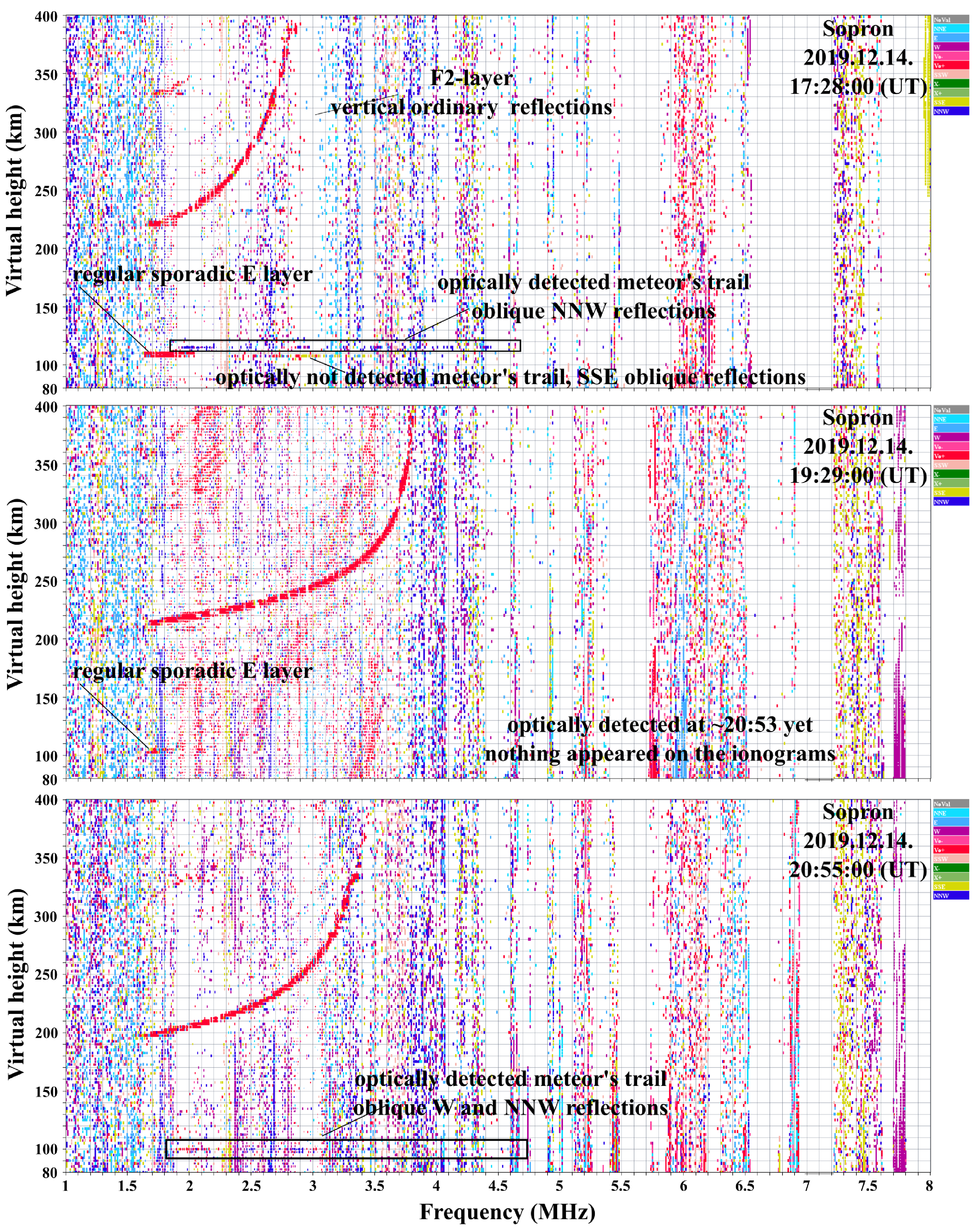,width=\textwidth}
\caption{\label{fig:2} Digisonde ionograms of the two simultaneously detected meteors (top and bottom
panels) and a non-detection example (middle panel). The top panel also shows the signal from another meteor
that was not detected optically. Virtual heights (in km)  are 
plotted on the vertical axis against radio frequency (in MHz) on the horizontal axis. These ionograms were
made applying the 2dB noise filtering algorithm instead of the default 6 dB in the 
Digisonde's software environment.}
\end{figure*}

\section{Instruments and data}

We used the Digisonde (DPS-4D Lowell Type Ionosonde), operating at Széchenyi István
Geophysical Observatory, Nagycenk, Hungary (SZIGO, 47.632°N, 16.718°E) for scanning the
ionosphere and detecting meteors during the 2019 Geminid shower (Dec 13 and 14, 2019) 
(see \cite{szarnya24} for more details 
and data\footnote{\tt https://data.mendeley.com/drafts/22n6tvjvv4}).

Additionally, two automated optical cameras were applied for simultaneous meteor detections.
One of them was a Watec 902H2 Ultimate camera having $122 \times 97$ deg$^2$ field-of-view and roughly
$+1$ limiting magnitude was operating at SZIGO.  The data were processed by the Metrec automatic 
meteor detection software (\cite{molau99}). The photometric accuracy was $\pm 0.5$ magnitude, 
according to the software description. 
Using a reference file of 281 bright stationary stars the software determines the right ascension 
and the declination along the meteor path as a function of time. It also identifies the parental meteor shower 
based on the measured path and apparent velocity (in degrees/s) of the meteor.

The other instrument was the 4-camera station of the Konkoly Meteor Observatory Network (KoMON), 
containing an ASI174MM video camera for triggering and a Canon 350D DSLR camera for imaging in each unit, 
having a $90 \times 270$ deg$^2$ combined field-of-view, operating at Gothard Astrophysical Observatory 
at Szombathely, Hungary (47.2578$^o$N, 16.6031$^o$E) (see \cite{deme1} for more details).

Although 79 meteors were detected by the optical camera at SZIGO (see \cite{szarnya24}), 
only 6 of them were
captured simultaneously by the KoMON system at Gothard. Among these, 2 meteors could also
be identified on the ionograms measured by the Digisonde at SZIGO on Dec 14, 2019.

Fig.\ref{fig:1} shows the optical frames of the two simultaneously detected meteors. 
Their basic measured properties are collected in Table~\ref{tab:1}.

Fig.\ref{fig:2} shows the ionograms of the two simultaneously detected meteors (top and bottom
panels). Frequency of the radio signal (in MHz) is plotted on the horizontal axis, while the
calculated virtual height (in kilometers) is shown on the vertical axis. The meteor signal
is marked by the black box in both panels. In the middle panel an ionogram is presented
that does not contain any meteor signal, although there was a contemporaneous meteor
event detected by the optical camera.

\section{Results}
The trajectory of the two simultaneous meteors in the atmosphere were determined by applying the method
of intersecting planes (\cite{vida20}). From the inferred trajectory the heights of the meteors at the moment of their appearance and disappearance ($H_{beg}$ and $H_{end}$, respectively) were calculated. These data are shown in
Table~\ref{tab:2}, together with the mean virtual heights ($H_{virt}$) measured on the ionograms provided by the
Digisonde. The virtual heights were then corrected for the elevation of the meteor at the beginning of
the optical detection as $H_{corr} = H_{virt} \times \sin({\rm ALT(beg)})$.

By comparing the 2nd and the 7th column in Table~\ref{tab:2} it is seen that the corrected heights of the Digisonde meteors are in very good agreement with the beginning height of the trajectory of the corresponding optical meteors as measured at SZIGO (the same site as that of the Digisonde). The difference between the two heights is less than the resolution of the Digisonde measurement, about 5 km. It is an important result regarding the interpretation of the Digisonde data: it seems that by taking into account the meteor elevation, measured by an optical camera, it is possible to convert the Digisonde virtual heights into true physical heights above the surface of the Earth. 

Moreover, we compared the inferred heights of the two simultaneous meteors with those expected for the Geminid shower (\cite{vida22}). Both of them are consistent with being Geminids, i.e. the beginning heights are in between 72 and 112 km. 

We plan to repeat this observing campaign in order to have more simultaneous meteors and confirm the results in the near future.  

\section{Conclusion}

The results of the present study are summarized as follows. 

\begin{itemize}

\item{the atmospheric heights of the meteor trails measured by the Digisonde, after the geometric correction
for meteor elevation are in very good agreement with the heights inferred from the
meteor trajectories reconstructed from the optical data;} 
\item{the difference between the two
measurements is within the vertical resolution limit of the Digisonde, about 5 km;}
\item{the simple geometric correction for off-zenith sources accounts for most of the distortion in the
virtual heights provided by the Digisonde;}
\item{the inferred optical meteor heights are also in good agreement with the expected atmospheric
heights for the Geminids, between 72 - 112 km.}
\end{itemize}

Acknowledgements: \\
The deployment and the operation of the Digisonde at SZIGO and the KoMON system was supported by the
"Cosmic Effects and Risks” GINOP 2.3.2-15-2016-00003 grant of the Hungarian Government, funded by the
European Union. \\
This work was supported by the project "Spatial and temporal variability of midlatitude sporadic E over Central
Europe” (KMP-2023/77). \\
The authors appreciate support from the bilateral project of the Czech Academy of Sciences and Hungarian
Academy of Sciences entitled "Multiinstrumental investigation of the midlatitude ionospheric variability” (n.
MTA-19-03 and NKM 2018-28) in facilitating scientific communication. \\
The contribution of VB was supported by Bolyai Fellowship (GD, No. BO/00461/21). The work of LD is supported by the HUN-REN SA-95/2021 project. Special thanks are due to Nagykanizsa Amateur Astronomers Association (especially Zsolt Perkó and Attila Gazdag), and the Bárdos Lajos Primary School, Fehérgyarmat (especially Zoltán Pásztor) for their kind contributions.

\bibliography{digisonde.bib}



\end{IMCpaper}

\end{document}